\documentclass[11pt,twoside]{article}
\usepackage{asp2010}

\resetcounters

\begin{document}

\title{Polarimetry of cool atmospheres: From the Sun to exoplanets}
\author{Svetlana V.\ Berdyugina
\affil{Kiepenheuer Institut f\"ur Sonnenphysik, D-79104 Freiburg, Germany}
}

\begin{abstract}
This is a review of a decade-long effort to develop novel tools for exploring magnetism in cold astrophysical media and to establish a new field of molecular spectropolarimetry since Berdyugina et al.~(2000). It is most directly applicable to the Sun, cool stars, substellar objects, planets and other minor bodies as well as interstellar and circumstellar matter. It is close to
being a mature field with developed theoretical tools poised to uncover new insights into the magnetic structures in cooler environments. Here I attempt a  broad description of the literature and present some recent exciting results. In particular, following my programmatic review at SPW3, I discuss advances in molecular magnetic diagnostics which are based on the modeling of about a dozen diatomic molecules with various electronic transitions and states, including the most challenging -- FeH. The applications stretch from sunspots to starspots, small-scale and turbulent solar magnetic fields, red and white dwarfs, and spin-offs such as polarimetry of protoplanetary disks and exoplanets. 
\end{abstract}

\section{Advances in molecular diagnostics}\label{sec:molec}

Interest in both the theory and experimental applications of the molecular Zeeman effect has had several ups and downs during the last hundred years. The peaks were around the 1930s, when the quantum mechanics was being established \citep[e.g.,][]{Kronig1928,Hill1929,Crawford1934}, in the 1960s, when more advanced experimental facilities became available, and after the year 2000, when we discovered this niche for solar and stellar astrophysics \citep{Berdyuginaetal2000mol}. Since then we have learned to reliably model polarization of about a dozen diatomic molecules with various electronic transitions and states which are listed in Table~\ref{tab:mol} together with corresponding references. The upper and lower parts of the Table contain transitions which are either in the Zeeman or Paschen-Back regimes, respectively, at kG fields or lower. The references are also divided into two parts: the first lists theoretical work relevant for a given transition and the second lists the papers with modeling of particular lines under various conditions.

A multitude of possible applications of molecular spectropolarimetry was first reviewed at the SPW3  \citep{Berdyuginaetal2003spw3} and included solar and stellar magnetic fields at various scales. In this paper I have tried to collect a rather complete list of most relevant references for the past decade and review the recent progress in these applications and also present spin-offs from these developments such as polarimetry of protoplanetary disks and exoplanets.

\begin{table}[!ht]
\caption{Diatomic molecules having transitions in the optical and near IR that
are present in solar and stellar spectra and are useful for measuring magnetic fields. }
\smallskip
{\small
\begin{center}
\begin{tabular}{llccc}
\tableline
\noalign{\smallskip}
       &\multicolumn{1}{c}{System}& $\lambda$ & \multicolumn{2}{c}{References} \\
\cline{4-5}  
       &                          &  [\AA]    &   Theory    &  Modeling   \\
\noalign{\smallskip}
\tableline
\noalign{\smallskip}
TiO    &$\gamma: A^3\Phi-X^3\Delta$          &  7054 (0,0) & 4 & 1,5,8,0   \\
       &$\gamma': B^3\Pi-X^3\Delta$          &  6224 (0,0) & 4 & 5   \\
       &$\alpha: C^3\Delta-X^3\Delta$        &  4950 (0,0) & 4 & 5   \\
       &$\delta: b^1\Pi-a^1\Delta$           &  8860 (0,0) & 4 & 5   \\
       &$\beta: c^1\Phi-a^1\Delta$           &  5597 (0,0) & 4 & 5   \\
\noalign{\smallskip}
OH     &$X^2\Pi$                             & 15328 (3,1) & 4,19 & 5   \\
\noalign{\smallskip}
FeH    &$F^4\Delta-X^4\Delta$                & 10062 (0,0) & 4,14 & 5,13,14  \\
       &                                     & 8715 (1,0)  & 4,14 & 0      \\
\noalign{\smallskip}
C$_2$  &$d^3\Pi-a^3\Pi$                      & 5165  (0,0) & 4,19 & 9,10,15,+   \\
\noalign{\smallskip}
\tableline
\noalign{\smallskip}
OH       &$A^2\Sigma-X^2\Pi$            & 2800--3400       & 4,6,11    &  3      \\
\noalign{\smallskip}
CN       &$A^2\Pi-X^2\Sigma$            & 7350--11580      & 4,6,11,19 &  2,18    \\
         &$B^2\Sigma-X^2\Sigma$         & 3490--4215       & 4,6,11    &  12      \\
\noalign{\smallskip}
MgH      &$A^2\Pi-X^2\Sigma^+$          & 4780--5620       & 4,6,11,19 &  1,6,9,+   \\
         &$B'^2\Sigma^+-X^2\Sigma^+$    & 5370--7590       & 4,6,11    &        \\
\noalign{\smallskip}
CaH      &$A^2\Pi-X^2\Sigma^+$          & 6680--7590       & 4,6,11    &  8,0    \\
         &$B^2\Sigma^+-X^2\Sigma^+$     & 6170--6870       & 4,6,11    &  0    \\
\noalign{\smallskip}
CH       &$A^2\Delta-X^2\Pi$            & 4150--4400       & 4,6,11,16 & 5,15,17\\
         &$B^2\Sigma^--X^2\Pi$          & 3870--4120       & 4,6,11    & 15   \\ 
\noalign{\smallskip}
SH       &$A^2\Sigma-X^2\Pi$            & 3250--3300      & 6,11      & 7 \\
\noalign{\smallskip}
\tableline
\end{tabular}
\end{center}
}
{\small
References:
1: \citet{Berdyuginaetal2000mol},
2: \citet{Berdyuginaetal2001hrmol},
3: \citet{BerdyuginaSolanki2001oh},
4: \citet{BerdyuginaSolanki2002mol1},
5: \citet{Berdyuginaetal2003mol2},
6: \citet{Berdyuginaetal2005mol3},
7: \citet{BerdyuginaLivingston2002sh},
8: \citet{Berdyuginaetal2006cah},
9: \citet{Berdyuginaetal2002hanle},
10: \citet{BerdyuginaFluri2004},
11: \citet{Shapiroetal2007hanlepbe},
12: \citet{Shapiroetal2007},
13: \citet{Aframetal2007},
14: \citet{Aframetal2008},
15: \citet{Berdyuginaetal2007prl}
16: \citet{Uitenbroeketal2004},
17: \citet{AsensioRamosetal2004},
18: \citet{AsensioRamosetal2005},
19: \citet{AsensioRamosTrujilloBueno2006},
+: see more references in Sect.~\ref{sec:turb},
0: this paper.
}
\label{tab:mol}
\end{table}

\section{3D structure of sunspots}\label{sec:suns}

Knowing the internal structure of sunspots allows us to gain insight on the energy transport in strong magnetic fields and, thus, on the processes inside the convection zone, where solar magnetic fields are generated and amplified before emerging at the surface on various scales. Spectropolarimetric studies of sunspots is a large field in solar physics, and there is already an immense amount of information collected during several decades \citep[e.g.][]{Solanki2003}. Nevertheless, there are still only a few studies on the 3D structure of sunspots which involve depth-dependent inversions of Stokes profiles of multiple lines. The situation becomes even more complicated in spectra of the coolest parts of the umbra where molecular lines become too prominent to be ignored. This brings however the advantage to investigate umbra with a high sensitivity. For instance, it was shown by \citet{Mathewetal2003} that simultaneous inversions of the infrared OH and \ion{Fe}{i} lines at 1.56\,$\mu$m extends the sunspot model to higher atmosphere layers and greatly constrains temperature, magnetic field, and flows. Simultaneous inversions of the \ion{Fe}{i} 5250\AA, TiO, and MgH lines holds promise to further improve average models of the sunspot umbra \citep{Aframetal2006spw4,Arnaudetal2007themis,Wenzeletal2010soho}.

\begin{figure}[!ht]
\centering
\includegraphics[width=0.27\linewidth]{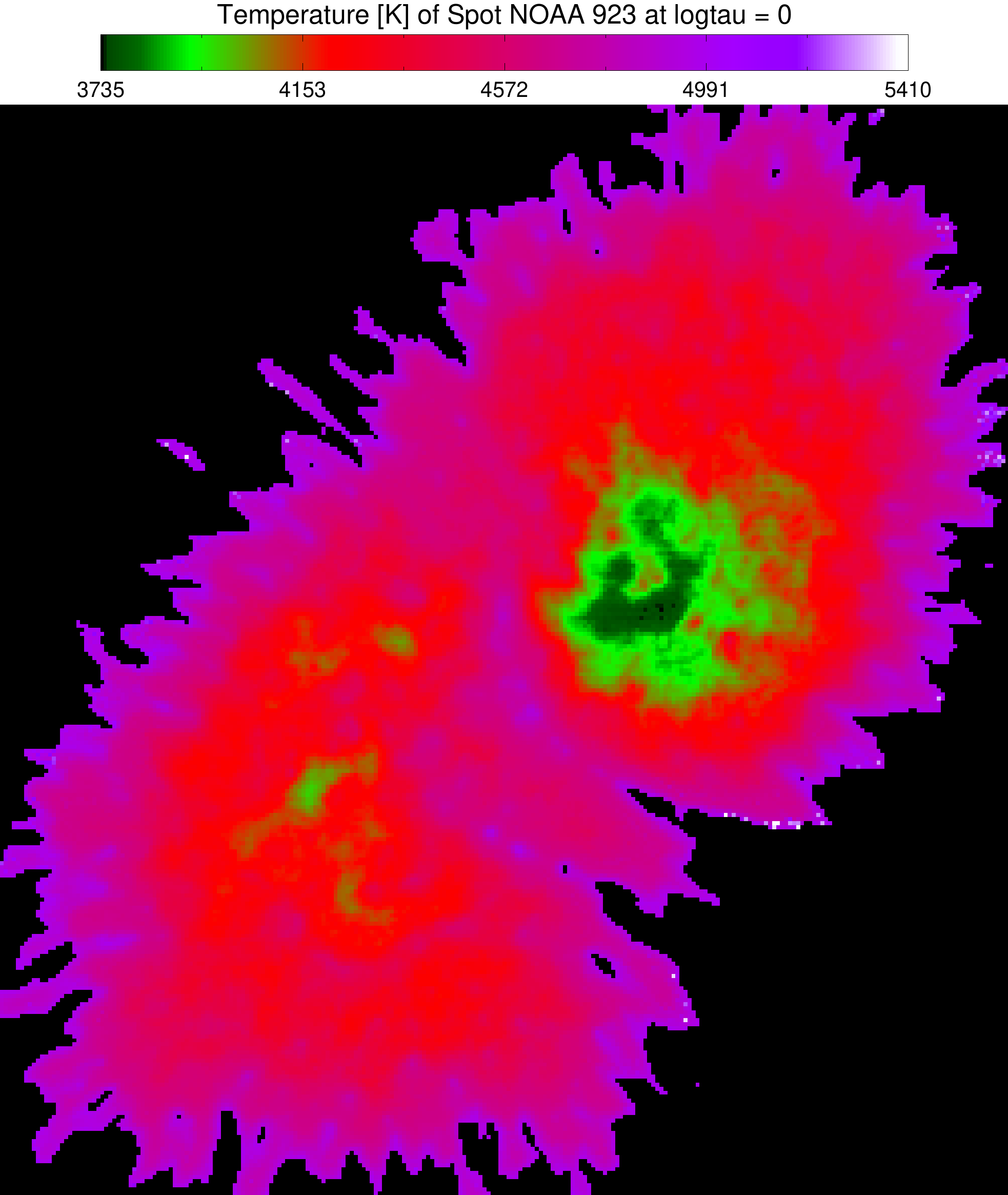}
\includegraphics[width=0.27\linewidth]{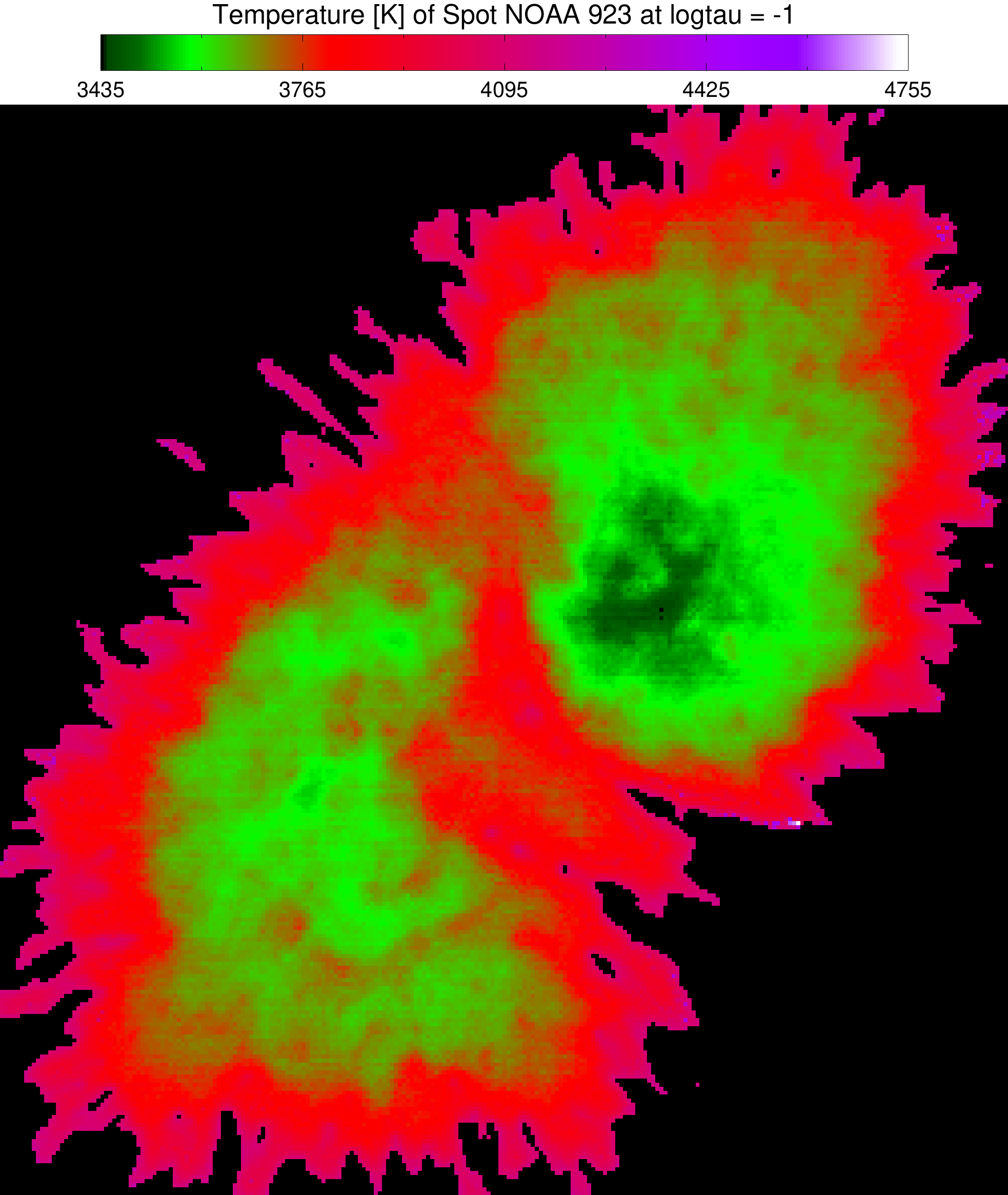}
\includegraphics[width=0.27\linewidth]{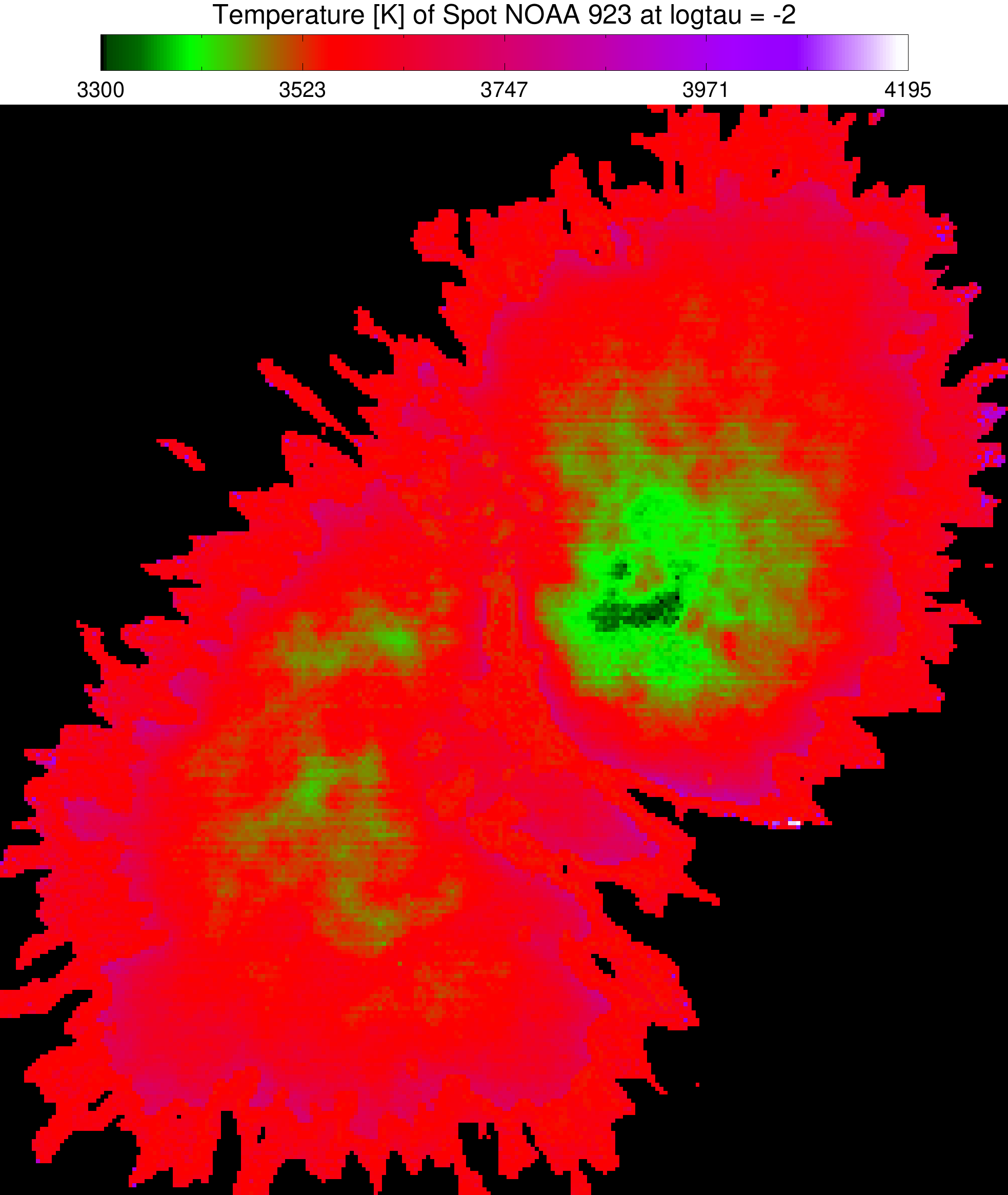}
\includegraphics[width=0.27\linewidth]{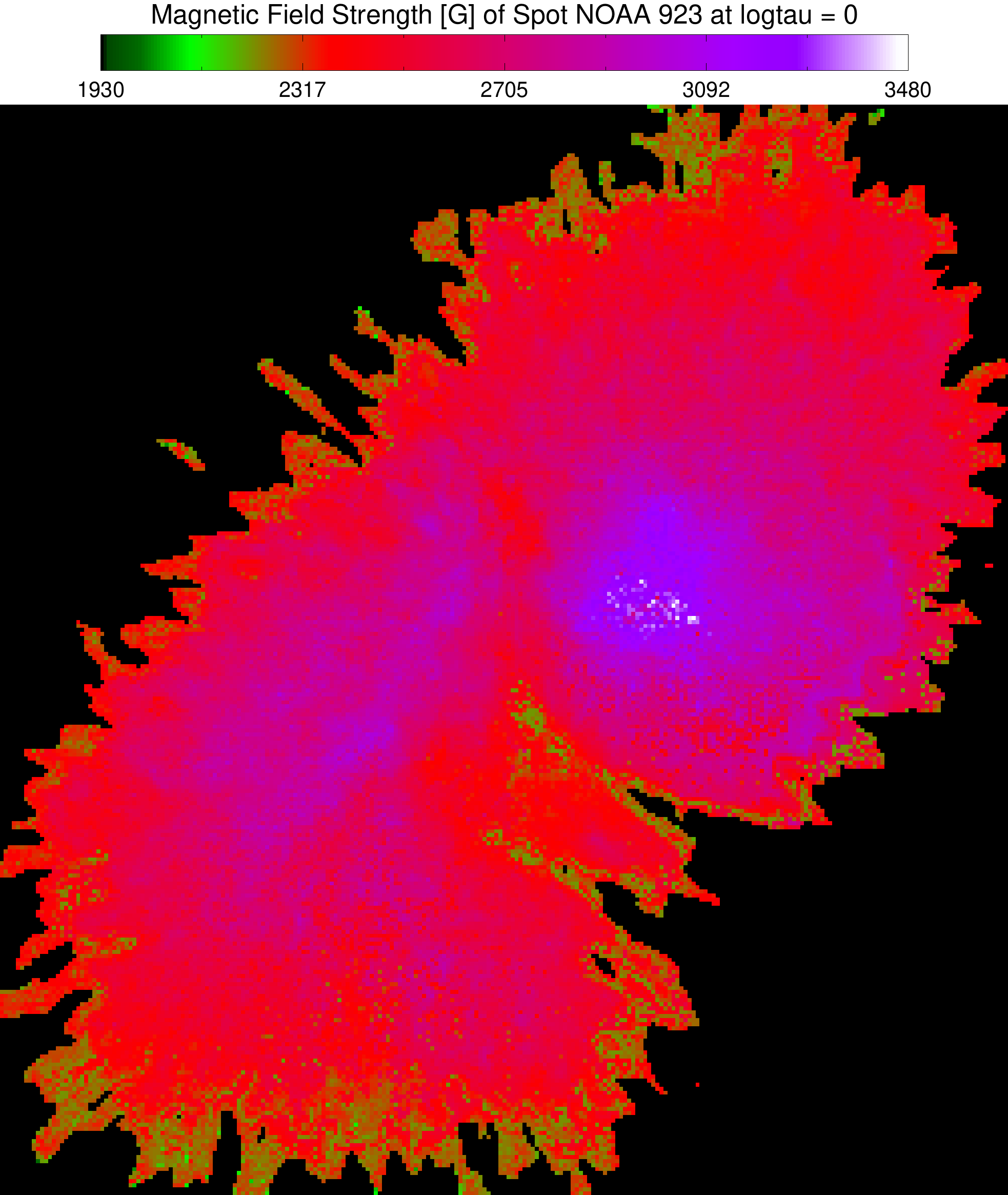}
\includegraphics[width=0.27\linewidth]{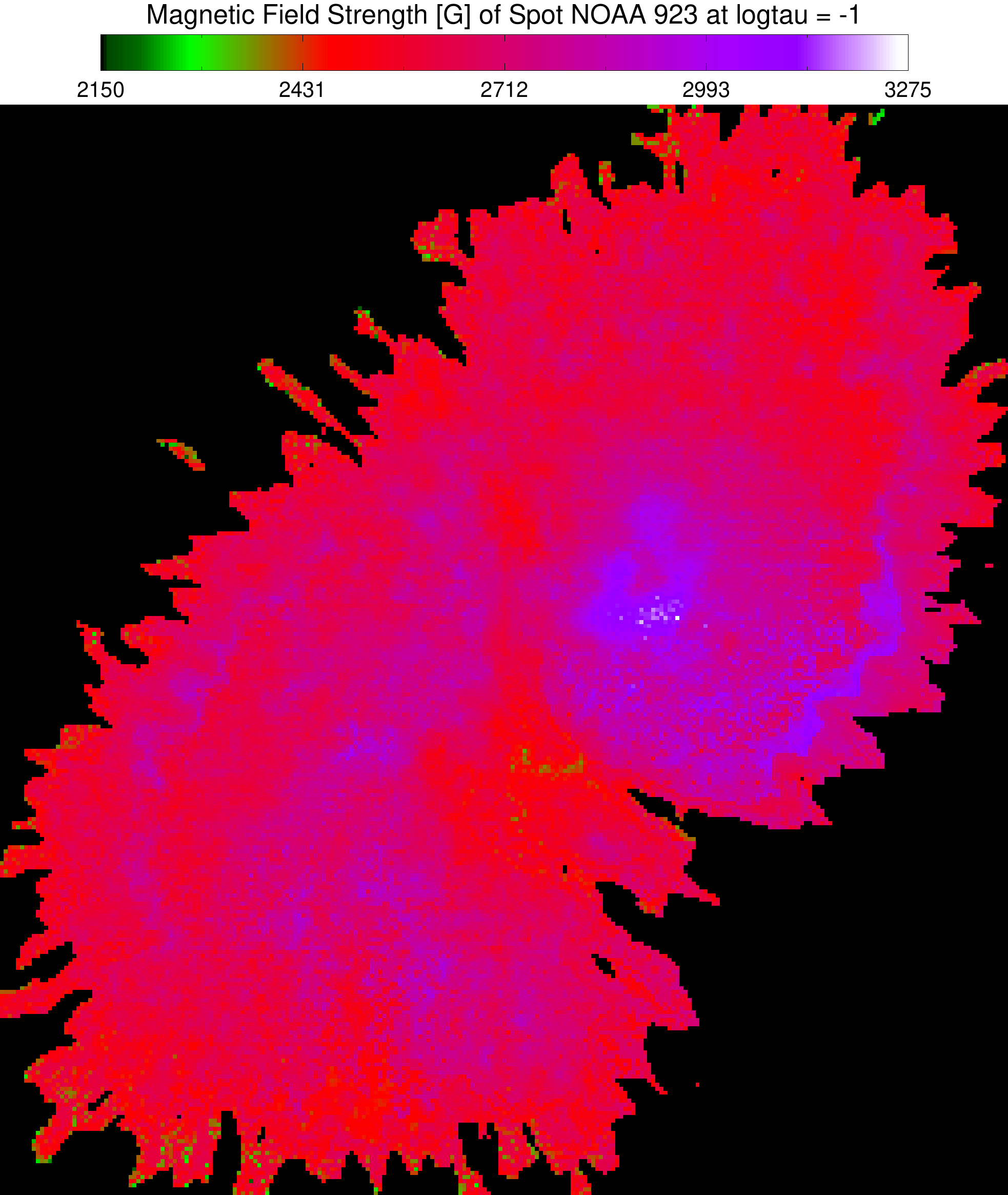}
\includegraphics[width=0.27\linewidth]{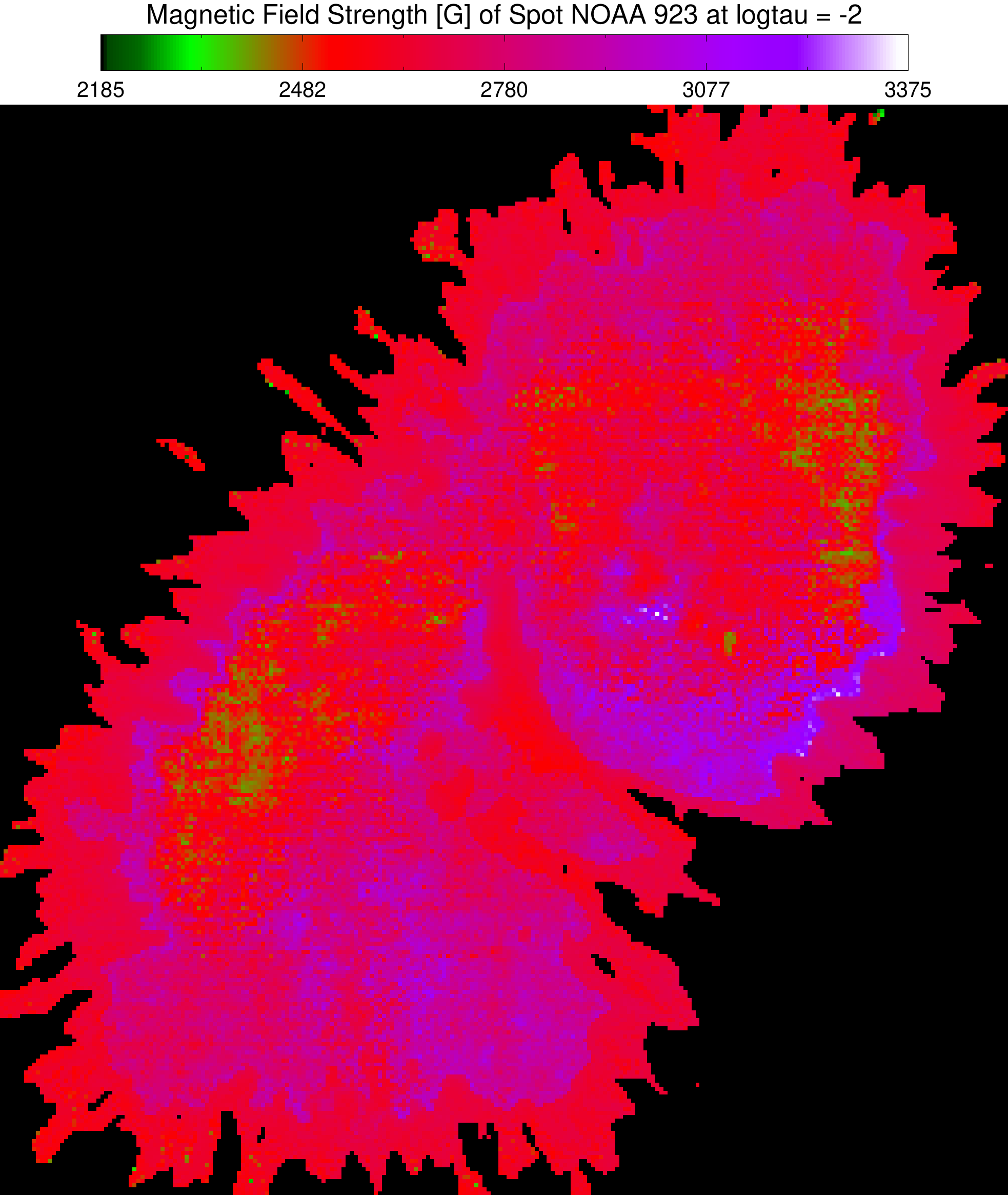}
\caption{Temperature (top row) and magnetic field (bottom row) distributions in the umbra of the sunspot NOAA 923 at different depths. Left to right: $\log\tau$= 0, -1, -2. The scaling is different for each panel. While the temperature drops significantly towards the top, the magnetic field strength does vary much.}
\label{fig:sunsTB}
\end{figure}

\begin{figure}[!ht]
\centering
\includegraphics[width=0.4\linewidth]{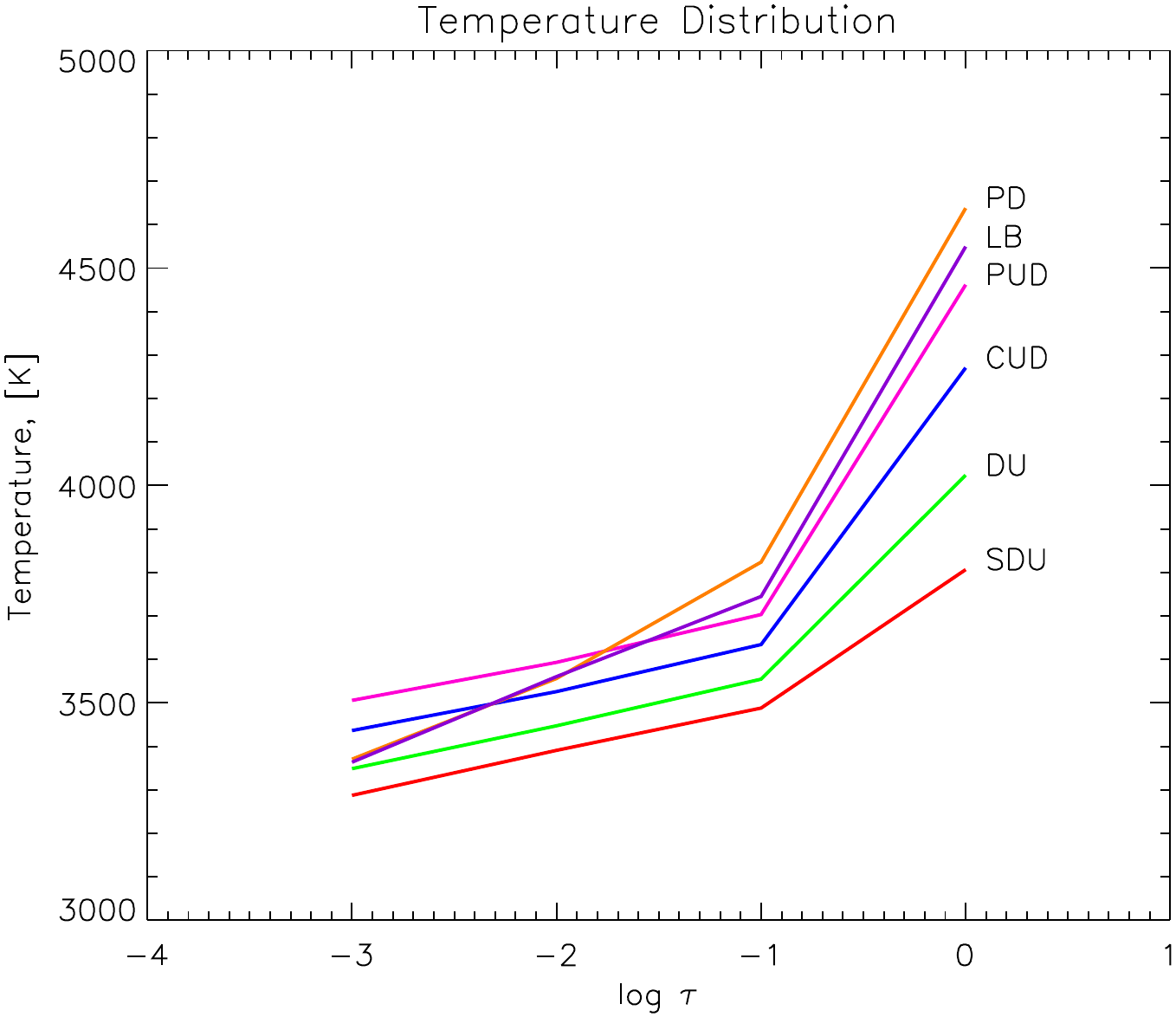}
\includegraphics[width=0.4\linewidth]{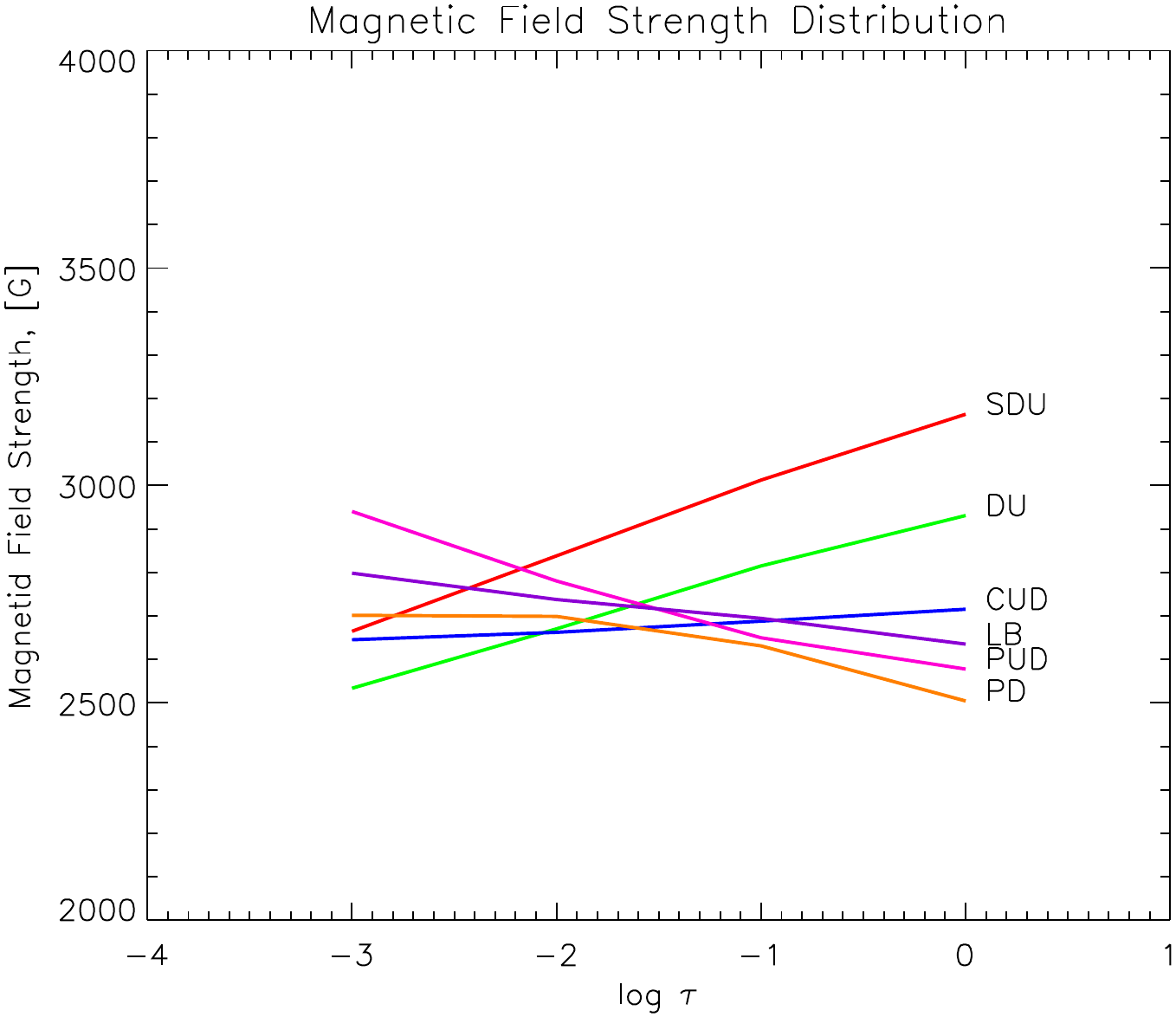}
\caption{Depth dependence of the temperature and magnetic field for different substructures of the sunspot umbra shown in Fig.~\ref{fig:sunsTB}. PD: penumbral dots, LB: light bridge, PUD: periferal umbral dots, CUD: central umbral dots, DU: dark umbra, SDU: darkest umbra.}
\label{fig:suns_mod}
\end{figure}

It was also found that accounting for contributions of CaH and TiO lines in inversions of the \ion{Fe}{i} 6302\AA\ region is important for obtaining realistic models of the sunspot umbra (in prep.). Figure~\ref{fig:sunsTB} shows our recent results obtained for the umbra of the sunspot NOAA 923 observed by Hinode from simultaneous inversions of atomic and molecular lines in the 6302\AA\ region. Thanks to the high spatial resolution of the the Hinode data, this provides a very detailed 3D structure of the umbra and allows us to extract average models of various substructures in the umbra (Fig.~\ref{fig:suns_mod}). This can be compared with the MHD models of sunspots by \citep{Rempeletal2009} and provide further constraints for the simulations. Furthermore, realistic model atmospheres of sunspots are important for modeling solar irradiance variations which are modulated by sunspots on short-time scales.

\section{Imaging solar magnetic fields in molecular bands}\label{sec:imag}

Imaging the solar surface in the G-band, where the main absorption is due to CH transitions, became a standard observing tool for studying the finest structure of small-scale magnetic fields and their dynamics \citep[e.g.,][]{BergerTitle2001}. It was identified that the dissociation process of the CH molecule in hotter magnetic environments is the reason for their high contrast in such images  \citep{Steineretal2001,SanchezAlmeidaetal2001,Schuessleretal2003,Shelyagetal2004}. 
\begin{figure}[!ht]
\centering
\includegraphics[clip,width=0.8\linewidth]{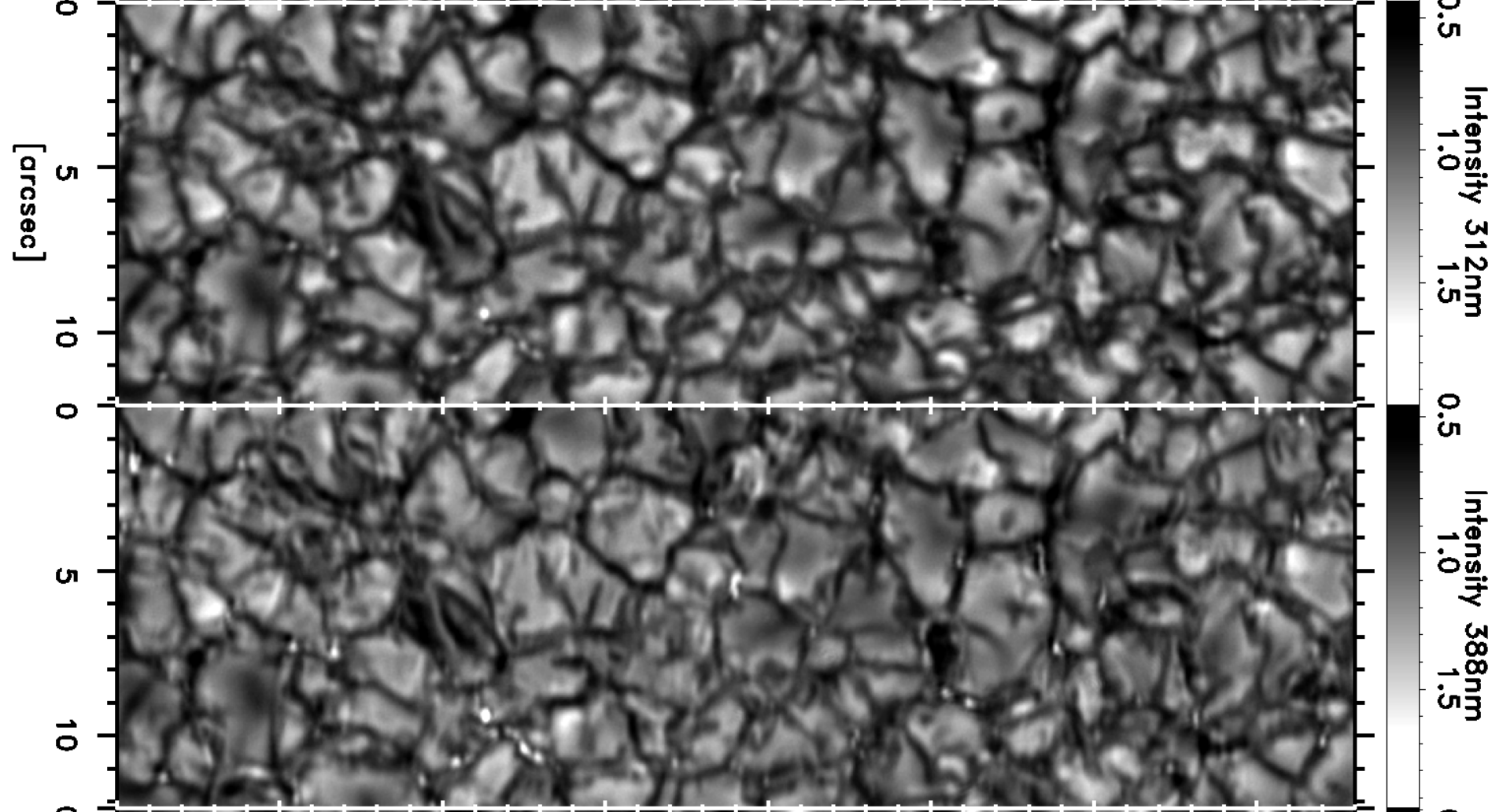}
\caption{Solar photosphere imaged in the UV OH (top) and CN (bottom) bands with the balloon-borne telescope Sunrise. From \citep{Hirzbergeretal2010}.
}
\label{fig:sunrise}
\end{figure}
Subsequently, we have shown that similar effects can be observed with other molecular bands, for instance in the violet systems of CN and OH \citep{Berdyuginaetal2003mol2}. Their dissociation equilibrium is similar to that of CH and one should expect a similar effect of a higher contrast in their bands. Indeed, ground-based imaging observations in the CN band at 3880\AA\ have confirmed our prediction that the contrast is even higher in the CN band as compared to the CH band \citep{Zakharovetal2005,UitenbroekTritschler2006,UitenbroekTritschler2007}. More recently, first imaging in the UV OH band were obtained with the balloon-borne Sunrise telescope \citep{Hirzbergeretal2010}. Theoretical interpretation and comparison with other bands will follow. 

\begin{figure}[t]
\centering
\includegraphics[clip,width=0.7\linewidth]{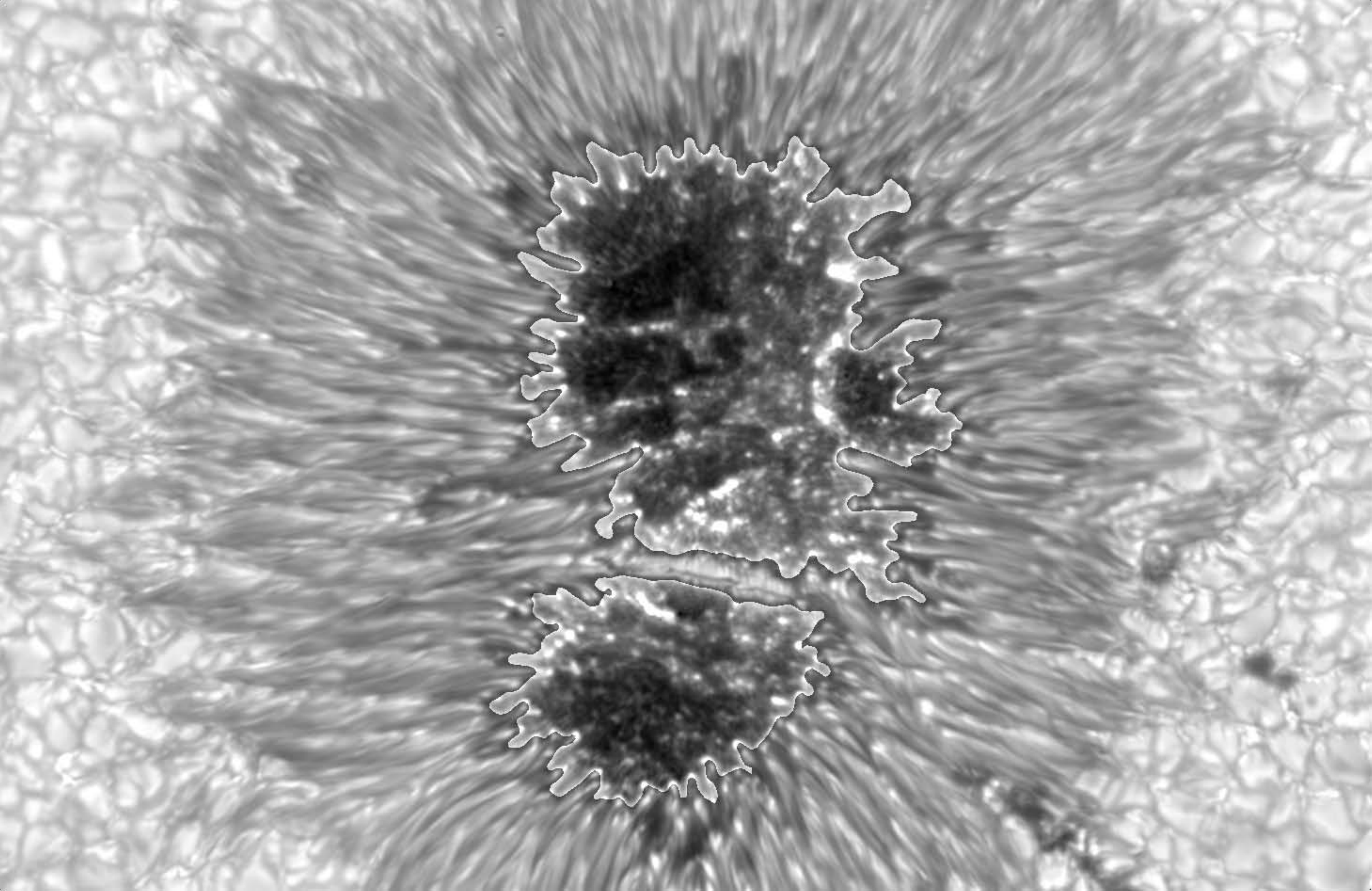}
\caption{The NOAA 10667 sunspot imaged in the TiO band by V.\ Zakharov during our coordinated observations with THEMIS and SST in September 2004. The umbra region is shown with an enhanced contrast to reveal an amazingly complex network of umbral dots.}
\label{fig:sunspotTiO}
\end{figure}

Molecular bands which are stronger at lower temperature are useful for imaging temperature inhomogeneities in sunspots. For instance, it was predicted \citep{Berdyuginaetal2003mol2} that umbral dots should be seen at higher contrast if observed in the TiO 7054\AA\ or UV OH band heads. Unfortunately there were no sunspots on the Sun during the Sunrise flight to verify this prediction, but ground-based observations at the SST, La Palma, in the TiO band were successful \citep{BergerBerdyugina2003}. Figure~\ref{fig:sunspotTiO} presents a sunspot imaged in the TiO band during our coordinated observing campaign with the SST and THEMIS (in prep.). It reveals a spectacular, never seen before, complexity in the umbra which is almost completely filled with umbral dots. This "web" of umbral dots appears to be a continuation of the penumbral structure but is deeply embedded in the umbra. This is in contrast to recent MHD simulations of sunspots by \citet{Rempeletal2009} where umbral dots are mere remnants of inhibited convection. Therefore, imaging and modeling of sunspots in the TiO band are important for understanding magneto-convection processes in the umbra. Furthermore, it is worth to mention that observing with the TiO filter apparently provides very sharp images of granulation and small-scale magnetic elements (but with a lower contrast than in G-band) and was even employed for recording Venus transit at SST (http://vt-2004.solarphysics.kva.se).
This filter was also successfully employed for first-light, highest resolution images at the new 1.7\,m telescope at the Big Bear Observatory \citep{Goodeetal2010}.

\section{Turbulent magnetic fields on the Sun}\label{sec:turb}

A study of turbulent magnetic fields on the Sun through scattering polarization and the Hanle effect is one of the major topics of the SPWs, and there are a number of contributions in this volume which address this topic. However, as they are devoted to atomic processes, I will try to cover here the key results obtained recently through molecular scattering polarization. The persistent questions concerning turbulent magnetic fields are (i) how strong are they on average, (ii) how are they distributed on the Sun, and (iii) whether they vary with time, e.g., with the solar cycle. It appears that the differential Hanle effect applied to molecular lines has the greatest potential to answer these questions. 

The first clues to puzzling molecular polarization in the second solar spectrum were provided by \citet{Berdyuginaetal2002hanle}. Even though in that work we did not discuss the effect of upper level lifetimes on the sensitivity to the Hanle effect, as was pointed out by \citet{TrujilloBueno2003} and \citet{Landi2003,Landi2007}, it appeared that the majority of lines within the same molecular band have similar lifetimes \citep{BerdyuginaFluri2004}, so this argument does not really clarify the question of why molecular polarization appeared nonvariable.

\begin{figure}[!ht]
\centering
\includegraphics[width=0.6\linewidth]{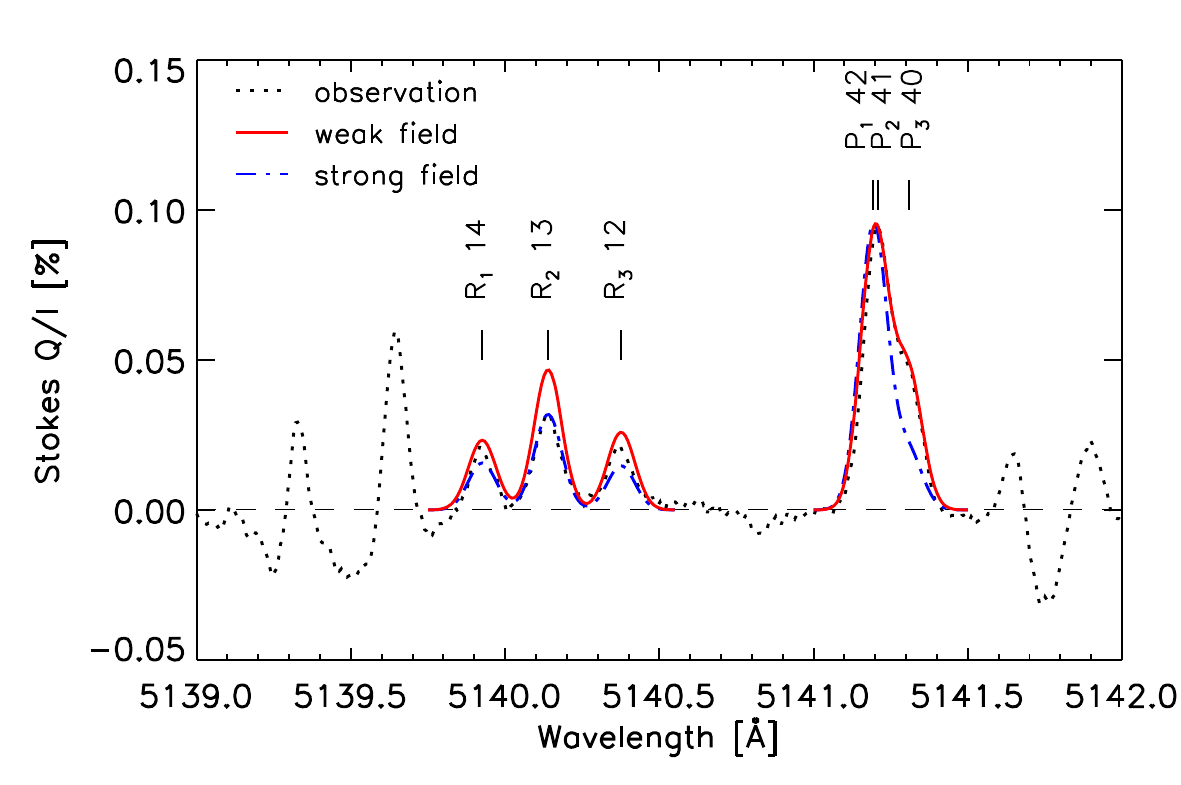}
\caption{Scattering polarization in two C$_2$ triplets. Two fits (solid and dash-dotted) to a synoptic observation (dotted) (averaged along the slit) are obtained using the differential Hanle model by \citet{BerdyuginaFluri2004}. The profiles shown with the solid line were calculated using a magnetic field of 7\,G and those shown with the dash-dotted line represent 90\,G. There is a clear indication that the field cannot exceed a few tens of Gauss but its value can only be determined if one assumes that the two triplets are differently affected by collisions. From \citet{Kleintetal2010c2obs}.}
\label{fig:c2}
\end{figure}

\begin{figure}[!ht]
\centering
\includegraphics[width=0.45\linewidth]{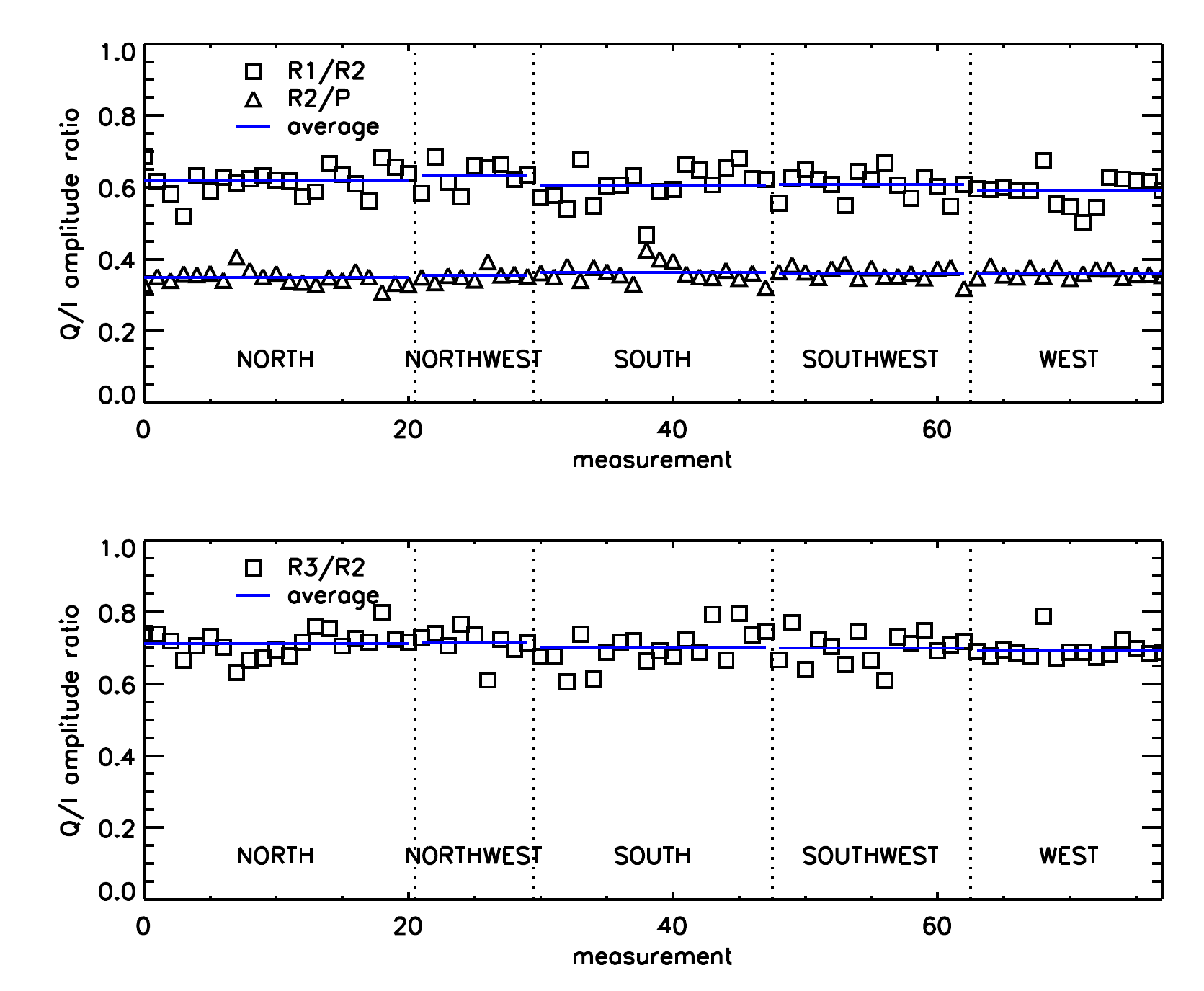}
\includegraphics[width=0.5\linewidth]{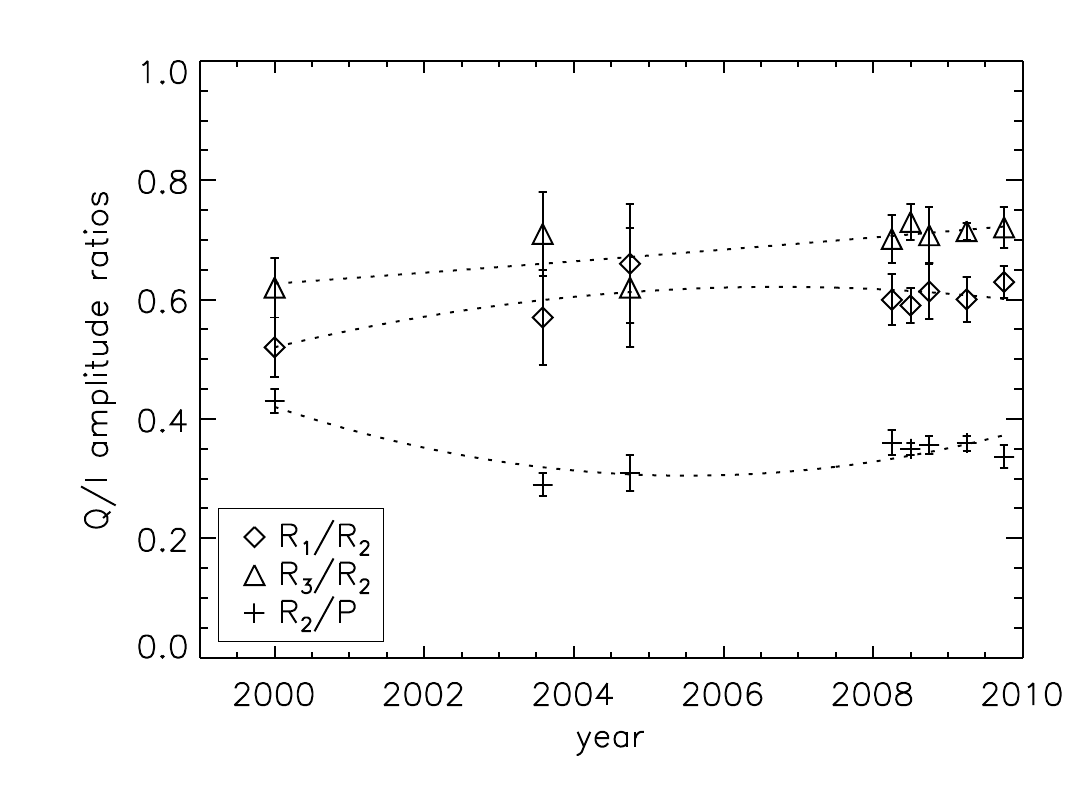}
\caption{
{\it Left}: $Q/I$ amplitude ratios for 78 synoptic measurements of the C$_2$ triplets shown in Fig.~\ref{fig:c2} sorted by heliographic position angle. The observed variations are within one standard deviation for all angles. This indicates a surprisingly homogeneous distribution of turbulent magnetic fields during the solar minimum of 2008--2009. 
{\it Right}: Amplitude ratio variations with time. A systematic decrease of R2/P and increase of R1/R2 and R3/R2 seem to occur as the solar activity declines. Second-order polynomial fits to the line ratios are shown with dotted lines. From \citet{Kleintetal2010c2obs}.
}
\label{fig:syn}
\end{figure}

Answering the first question above is a nontrivial task. The strength of the entangled magnetic field strongly depends on its assumed distribution within the resolution element on the Sun as well as on the model atmosphere used \citep[e.g.,][]{TrujilloBuenoetal2004}. Most results so far were obtained assuming an isotropic distribution, which provides a simple analytical tool to infer the field strength \citep{Stenflo1982}. When this is applied to lines with similar thermodynamical properties but with different magnetic sensitivities, the differential Hanle effect provides the most model independent field strength. At the SPW3 \citet{TrujilloBueno2003} argued that the observed difference in scattering amplitudes within partly resolved triplets of C$_2$ transitions (such as the P-triplet at 5141\AA\ in Fig.~\ref{fig:c2}) in the second solar spectrum are due to the Hanle effect. However, it could also be explained by a simple blending effect of two unresolved triplet components as compared to a partly resolved one. In search for the truth, we have carefully investigated the atlas of \citet{Gandorfer2000} and found a fully resolved and practically unblended R-triplet at 5140\AA\ \citep[][see also Fig.~\ref{fig:c2}]{BerdyuginaFluri2004}. This triplet was clearly affected by the Hanle effect and not the way it was argued before. However, even in this clear case the interpretation was hindered by unknown parameters, such as Einstein coefficients, which are not trivial in the case of C$_2$ transitions, and collision rates, which are not known with the necessary accuracy \citep[e.g.,][]{AsensioRamosTrujilloBueno2005,Bommieretal2006,Kleintetal2010c2obs}. These issues are not still clear and need further investigation. However, to summarize, it appears that many studies of the Hanle effect in molecular lines deduce a relatively low turbulent magnetic field strengths of 5--30\,G \citep{Faurobertetal2001,FaurobertArnaud2003,BerdyuginaFluri2004,Derouichetal2006,TrujilloBuenoetal2006,Bommieretal2006,Shapiroetal2007,Kleintetal2010c2obs}, which may be an indication for a depth-dependent effect. It is clear however, as illustrated in Fig.~\ref{fig:c2}, that the field strength in the lower atmosphere cannot exceed a few tens of Gauss, in contrast to conclusions by, e.g., \citet{TrujilloBuenoetal2004}.

In such a situation the second and the third questions cannot be answered without a systematic and homogeneous study. That is why we have initiated a synoptic program at IRSOL for measuring during several years (hopefully through at least one solar cycle) the scattering polarization in the C$_2$ triplet and a few atomic lines. The results for the first two years covering the solar minimum in 2008--2009 were reported by \citet{Kleintetal2010c2obs,Kleintetal2011c2mod}. Figure~\ref{fig:syn} illustrates that turbulent fields were evenly distributed over the Sun during this solar minimum. The field strength was on average at least 4.7$\pm$0.2\,G (without collisions taken into account) or 7.41$\pm$0.09\,G (with collisions and a NLTE analysis). It did not vary during the last two years, but there could be a very small variation of the turbulent field strength (3$\sigma$-limit) since the solar maximum in 2000 \citep[based on data of][]{Gandorfer2000}. This result is however very sensitive to the available statistics, so we need to continue the synoptic program to clarify whether turbulent magnetic fields are generated by a global dynamo responsible for the solar cycle or a local dynamo which may not vary at all on the cycle time scale. It is worth to mention that such a program would be a rather minor expense for large solar telescopes such as ATST \citep[see][]{Kleintetal2010syn}.


\section{3D Structure of Starspots}\label{sec:star}

Sunspots are the primary evidence of solar activity, and they harbour the strongest field and the coldest plasma on the solar surface. Such a phenomenon is common among stars possessing convective envelopes, where magnetic fields are believed to be generated \citep{Berdyugina2005lr}. However, magnetic fields were never measured inside starspots, and their internal structure remained unknown \citep[e.g.,][]{Berdyugina2009,Strassmeier2009,DonatiLandsreet2009}. Studying this phenomenon is especially interesting for cool dwarfs, which are known to show strong magnetic activity and whose dynamo is apparently different from that of more massive cool stars, such as the Sun. The process of generation of strong magnetic fields in objects with very deep convection zones or even fully convective is not yet understood. We have developed a unique approach to probe their atmospheres and magnetospheres with polarimetry and reveal a 3D structure of starspots.

The prospects of molecular spectropolarimetry for exploring magnetic interiors of starspots were foreseen by \citet{Berdyugina2002an}, but adequate observational facilities were not available at that time. Once the high-resolution spectropolarimeter ESPaDOnS \citep{Donatietal2006} became available at the CFHT, our team obtained first polarimetric measurements of starspots on M and K dwarfs. We have detected circular polarization in many lines, including the first detection of polarization in TiO, CaH, and FeH transitions of 0.5--1\% with the noise level of 0.1\% \citep{Berdyuginaetal2006spw4starspots} and in chromospheric emission lines \citep{Berdyuginaetal2008cs14}. Here we summarize first results inferred from these data (in prep.) for an active red dwarf AU~Mic, M1Ve, which is the brightest X-ray source among such stars.  

\begin{figure}[!ht]
\centering
\includegraphics[width=0.96\linewidth]{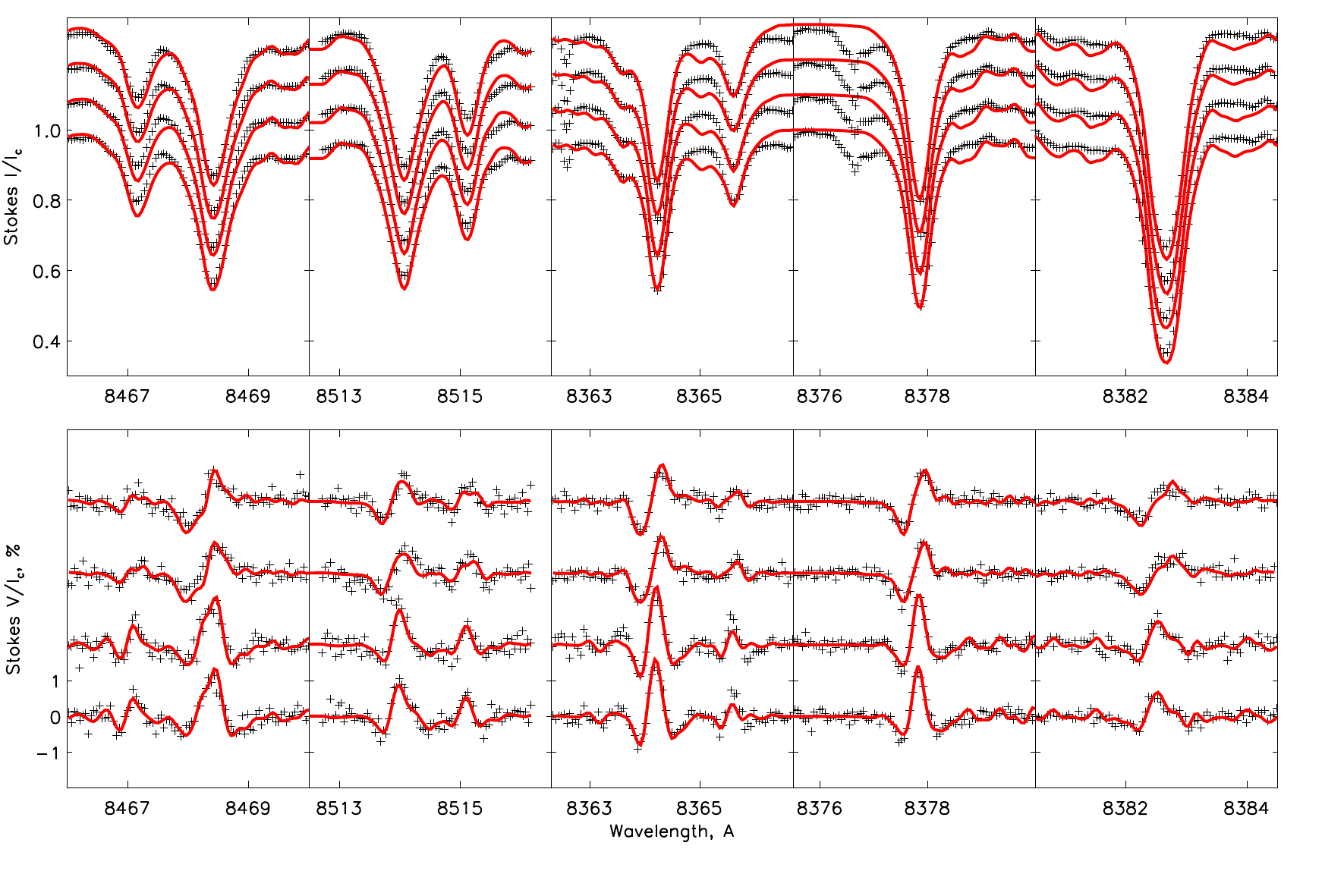}
\caption{Stokes profiles of \ion{Fe}{i} (two left panels)) and \ion{Ti}{i} (three right panels) lines in the spectra of AU~Mic. Data (crosses) for four rotational phases recorded during two consecutive nights are shifted in vertical for clarity. Model fits (red line) correspond to the images shown in Fig.~\ref{fig:aumic_depth}.
}
\label{fig:aumic_atoms}
\end{figure}

\begin{figure}[!ht]
\centering
\includegraphics[width=0.32\linewidth]{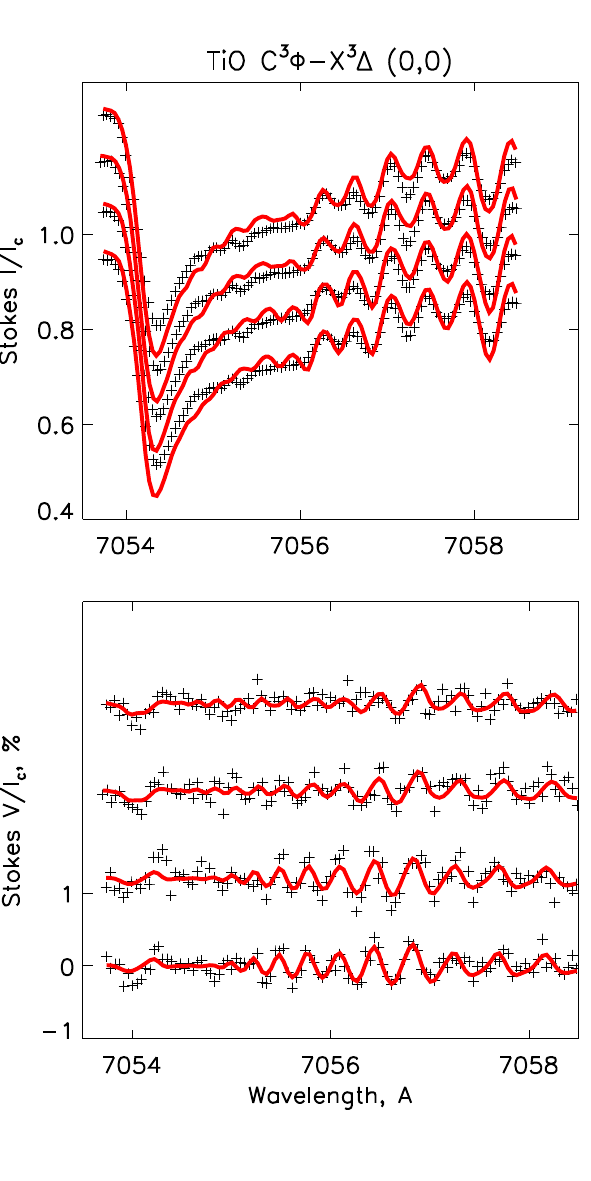}
\includegraphics[width=0.32\linewidth]{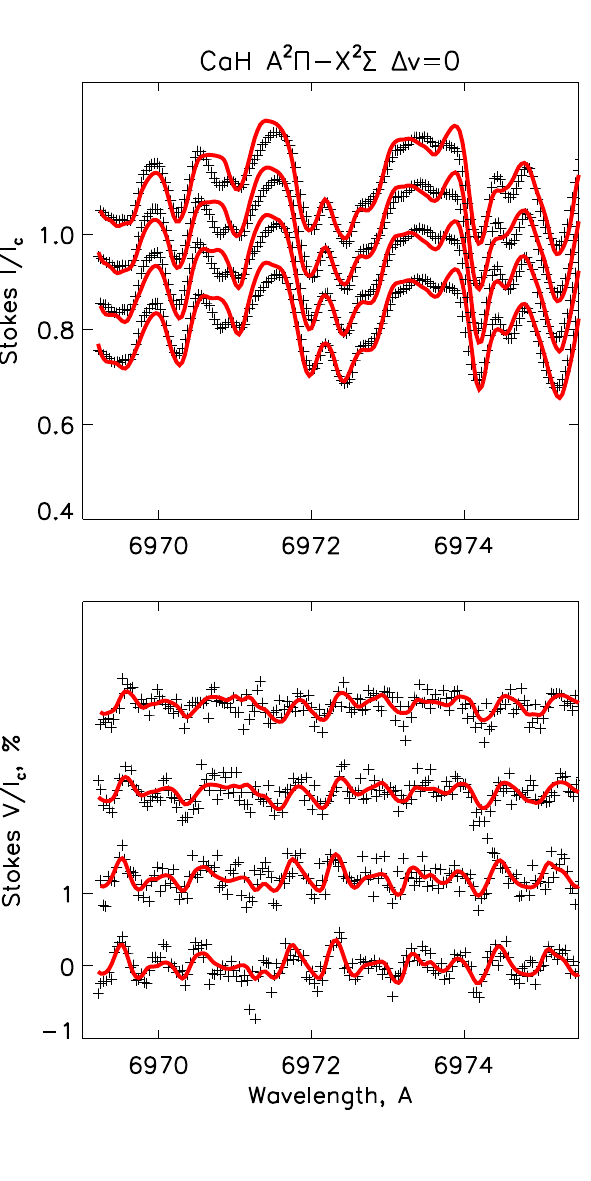}
\includegraphics[width=0.32\linewidth]{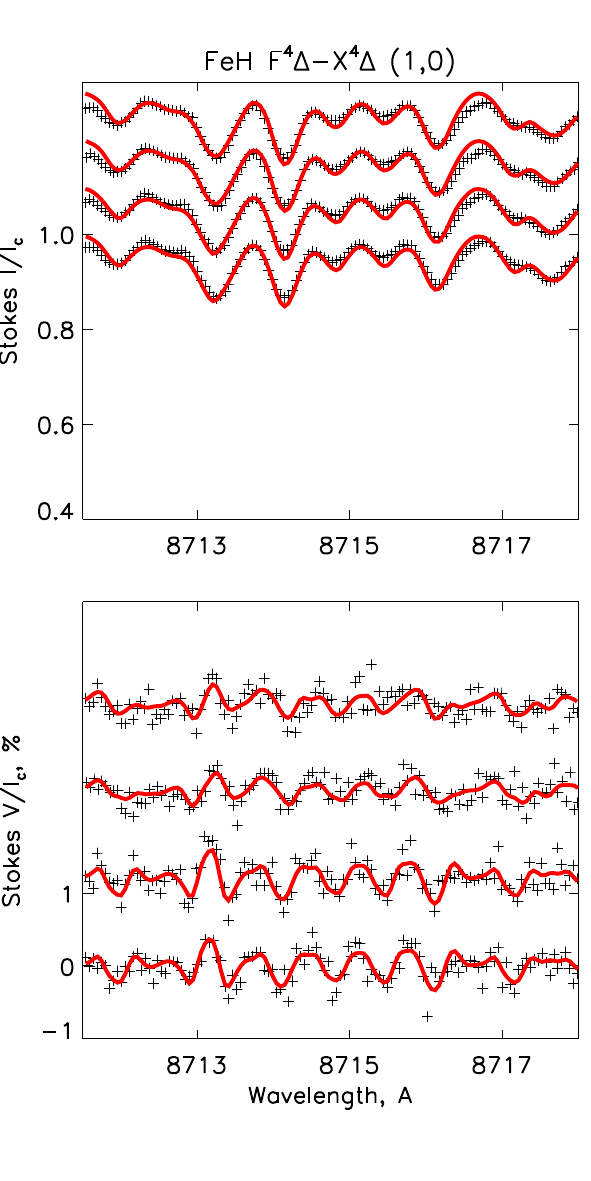}
\caption{The same as Fig.~\ref{fig:aumic_atoms} for molecular lines.}
\label{fig:aumic_molec}
\end{figure}

\begin{figure}[!ht]
\centering
\includegraphics[width=0.8\linewidth]{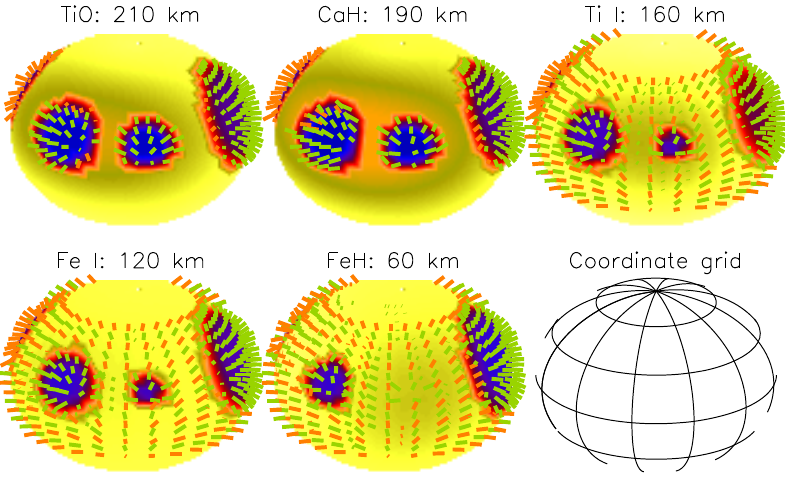}
\caption{Starspots on AU Mic at different heights above the photosphere obtained from different lines. Darker regions are cooler (scaling is individual for each image). Green and orange sticks depict magnetic vectors of away and towards the observer, respectively. They are scaled to the unit length corresponding to 5\,kG.}
\label{fig:aumic_depth}
\end{figure}

We selected several lines of \ion{Fe}{i} and \ion{Ti}{i} and many of TiO, CaH and FeH with prominent Stokes $V$ profiles (Figs.~\ref{fig:aumic_atoms} \&~\ref{fig:aumic_molec} ). These lines cover a wide range of temperature and magnetic sensitivity and also effectively form at different heights and locations in the stellar atmosphere. This provides a unique opportunity for probing directly the interior of starspots, basically independent on the distance to the star or its size, i.e.\ clearly beyond current direct spatial resolution on stellar surfaces. Since the observations of AU Mic were obtained during two consecutive nights, which roughly covers half the stellar rotation period, we can only analyse spots visible during that time. We applied our new Zeeman-Doppler imaging (ZDI) inversion technique \citep{Sennhauseretal2009spw5} and Stokes parameter synthesis \citep{Berdyuginaetal2003mol2} to individual species forming at different heights in the range of 60--210\,km above the photosphere. Due to lack of spatial information in the line profiles (because of the slow stellar rotation, $v\sin i$=10 km/s), shapes of spots remained uncertain and for simplicity were initially assumed to be circular and homogeneous, which was subsequently refined by inversions until the best fit to the data was achieved. The five obtained surface images are shown in Fig.~\ref{fig:aumic_depth}. 

\begin{figure}[t]
\centering
\includegraphics{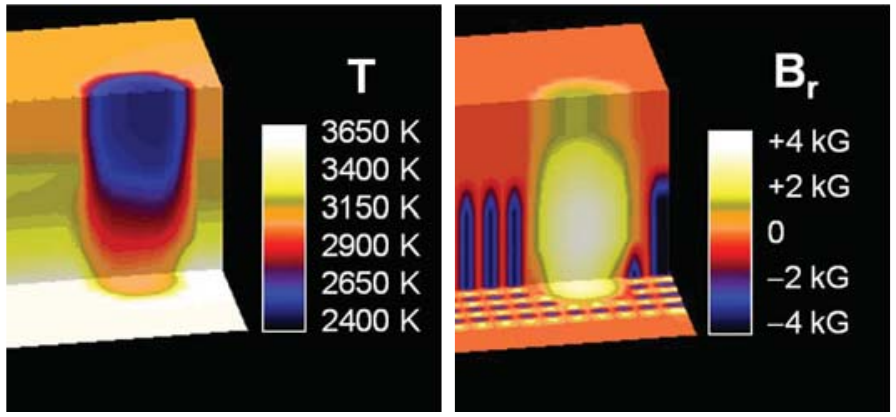}
\caption{3D structure of a starspot on AU~Mic (vertical cut through the atmosphere). The temperature and radial magnetic field decrease with height, while the spot area increases. An entangled field of 3.5\,kG (assumed to be radial) fills space between spots in the lower photosphere.}
\label{fig:aumic_3D}
\end{figure}

Spots on AU~Mic were found in a near-equatorial zone (their exact latitude is uncertain because of the unknown inclination angle of the stellar rotation axis). This is similar to the Sun but in contrast to rapidly-rotating active stars where Coriolis force brings spots closer to the rotational axis, i.e. poles. Even though the spot shapes are unknown, their area and longitudes are well constrained by the data. We identify four spots, one of positive and three of negative polarity (Fig.~\ref{fig:aumic_depth}). They are 500--700\,K cooler than the photosphere and harbor a maximum magnetic field of 5.3\,kG. The smallest spot is comparable in absolute dimensions (70\,Mm) with very large sunspots (60\,Mm), and it practically disappears at the bottom of the atmosphere. The vertical temperature and magnetic field distributions inside one starspot are shown in Fig.~\ref{fig:aumic_3D}. 

The spot temperature decreases with height from 3100\,K to 2400\,K ($\pm$100\,K), with the gradient $dT/dh$$\sim$5\,K\,km$^{-1}$ (Fig.~\ref{fig:aumic_3D}), which is significantly larger than that in sunspots. The magnetic field strength (modulus) in spots reaches more than 5\,kG, which is much stronger than the equipartition level. Its radial component (which was found to be dominant in spots) decreases with height from 4.0\,kG to 2.3\,kG ($\pm$0.5\,kG). The gradient $dB_{\rm r}/dh$$\sim$10\,G\,km$^{-1}$ is an order of the magnitude larger than that in sunspots. To exclude the influence of the density scale height, we consider the gradient $dB_{\rm r}/dT$, which is about 2\,G\,K$^{-1}$ in AU~Mic spots and about 1\,G\,K$^{-1}$ in sunspots, and they are both significantly larger than the adiabatic gradient. It appears that this difference occurs due to decoupling of the magnetic field from the ambient gas, whose ionization degree is much lower on AU~Mic than on the Sun. This goes well along with the fact that the spot area on AU~Mic increases with height, i.e.\ the field fans out and is loosely confined by the gas.

We find also evidence for the second magnetic component in the AU~Mic atmosphere: a mix-polarity field of 3.5$\pm$0.5\,kG covering about half of the photosphere outside cool spots. This field is apparently confined to the lower photosphere or perhaps weakens with height. It appears analogous to a solar network magnetic field. Its topology is unknown and for simplicity assumed to be radial. This result clarifies high magnetic filling factors deduced from measurements of Zeeman-broadened profiles, including FeH, in active M dwarfs \citep{Johns-KrullValenti1996,Valentietal2001,ReinersBasri2006}. With our advanced quantum-mechanical modeling of FeH transitions, the accuracy and reliability of such measurements has significantly improved \citep{Aframetal2009spw5mdw,Aframetal2011}.

Less active targets show however polarization signals significantly smaller than 1\%, while current observational facilities limit the accuracy per wavelength to 0.1\%. Therefore, for such targets it is not possible to detect signals in individual lines, and multi-line techniques such as LSD \citep{Donatietal1997} are commonly used to increase the signal-to-noise ratio (SNR). However, limitations of the LSD are now well recognized \citep[e.g.,][]{Berdyugina2009}, and new techniques emerge to advance our ability to measure and interpret Stokes profiles, such as PCA \citep{MartinezGonzalez2008}, NDD \citep{Sennhauseretal2009}, and ZCD \citep{SennhauserBerdyugina2010zcd}. The latter two are particularly powerful as they take into account nonlinear effects in blending spectral lines and their Zeeman components for both atomic and molecular transitions. At the moment the ZCD is the most sensitive technique for detecting very weak magnetic fields, such as about 0.5\,G on the K giant Arcturus \citep{SennhauserBerdyugina2010aboo}. When combined with the ZDI, it will open new opportunities to explore magnetic spots in many objects across the HR-diagram.

\section{White Dwarfs}\label{sec:wd}

Applications of molecular spectropolarimetry extend even to cool white dwarfs (WDs), where this technique is the only hope to clarify the frequency of magnetic objects and learn about their evolution. For instance, a small group of cool, helium-rich WDs (DQ type, T$\le$7000\,K) show only carbon-based molecular bands, mainly C$_2$ bands \citep[e.g.,][]{Dufouretal2005}, which are therefore the only indicators of their magnetic fields. It is believed that the majority of these stars are not magnetic,  since no polarization was reported in C$_2$ until now. However, a famous representative of this group, G99-37, for more than four decades having been a unique WD with both CH and C$_2$ molecular bands, is clearly magnetic, as indicated by prominent broad-band polarization in the CH bands discovered by \citet{AngelLandstreet1974}. This is due to magnetic dichroism appearing in molecular bands in the presence of a strong magnetic field, which perturbs the internal structure of the molecule and results in net polarization due to the Paschen-Back effect (PBE). Recently we analyzed new spectropolarimetric observations of G99-37 \citep{Berdyuginaetal2007prl} using our approach to the molecular PBE \citep{Berdyuginaetal2005mol3}.  In addition to
previously known molecular bands of the C$_2$ Swan and CH $A-X$ systems we found also a firm evidence for the violet CH $B-X$ bands at 390\,nm and C$_2$ Deslandres-d'Azambuja bands at 360\,nm. Combining the polarimetric observations with our model calculations, we deduced a dipole magnetic field of 7.5$\pm$0.5\,MG with the positive pole pointing towards the Earth. 

\begin{figure}[t]
\centering
\includegraphics[width=\linewidth]{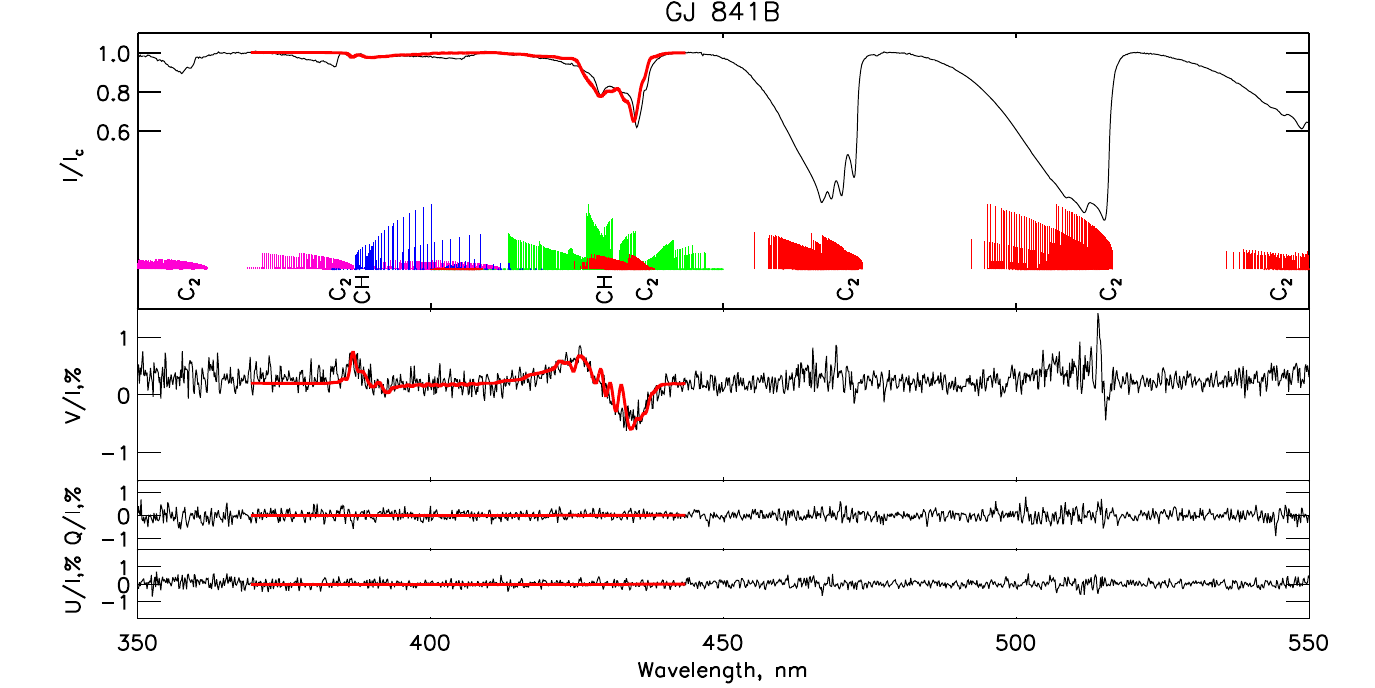}
\caption{Observed intensity distribution, circular and linear polarization of the white dwarf GJ 841B -- the second WD with CH bands (thin line). The CH and C$_2$ bands are identified. Synthetic spectra for a longitudinal field of 1.3\,MG directed to the observer shown by thick red lines.}
\label{fig:wd}
\end{figure}

Inspired by this result, we started a spectropolarimetric survey of cool DQ-type WDs. This has recently resulted in discovery of the second WD with polarized CH bands, in both $A-X$ and $B-X$ systems! The CH bands in the intensity spectrum of GJ 841B are completely masked by strong C$_2$ absorption, and only polarimetric measurements of high accuracy could reveal the CH bands and the magnetic nature of the WD \citep{Vornanenetal2010}. By modeling four Stokes profiles in the CH bands we determined the magnetic field strength of 1.3$\pm$0.5\,MG, which is largely longitudinal (Fig.~\ref{fig:wd}). It is interesting that the circular polarization is of the same sign as in G99-37, and the continuum polarization of +0.2\% is also present. In addition, GJ 841B is the first white dwarf to show signatures of weak polarization in the C$_2$ Swan bands. The discovery of the second WD with polarized CH and C$_2$ bands holds promise that there are more such objects out there. The developed technique of molecular magnetic dichroism has proved itself to be an excellent tool for studying magnetic fields on cool and faint magnetic objects.

\section{Exoplanets and Protoplanetary Disks}\label{sec:plan}

Two interesting spin-offs from our developments for scattering polarization in the solar atmosphere are polarimetric studies of exoplanets and of protoplanetary disks. Here we summarize some of the major results obtained for exoplanets in recent years. The development of a new diagnostic technique based on radiative pumping of absorbers to explain ubiquitous absorption line polarization in embedded stars and to explore inner parts of protoplanetary disks has been pioneered by \citet{Kuhnetal2007} and is reviewed in these proceedings by \citet{Kuhnetal2011}.

\begin{figure}[t]
\centering
\includegraphics{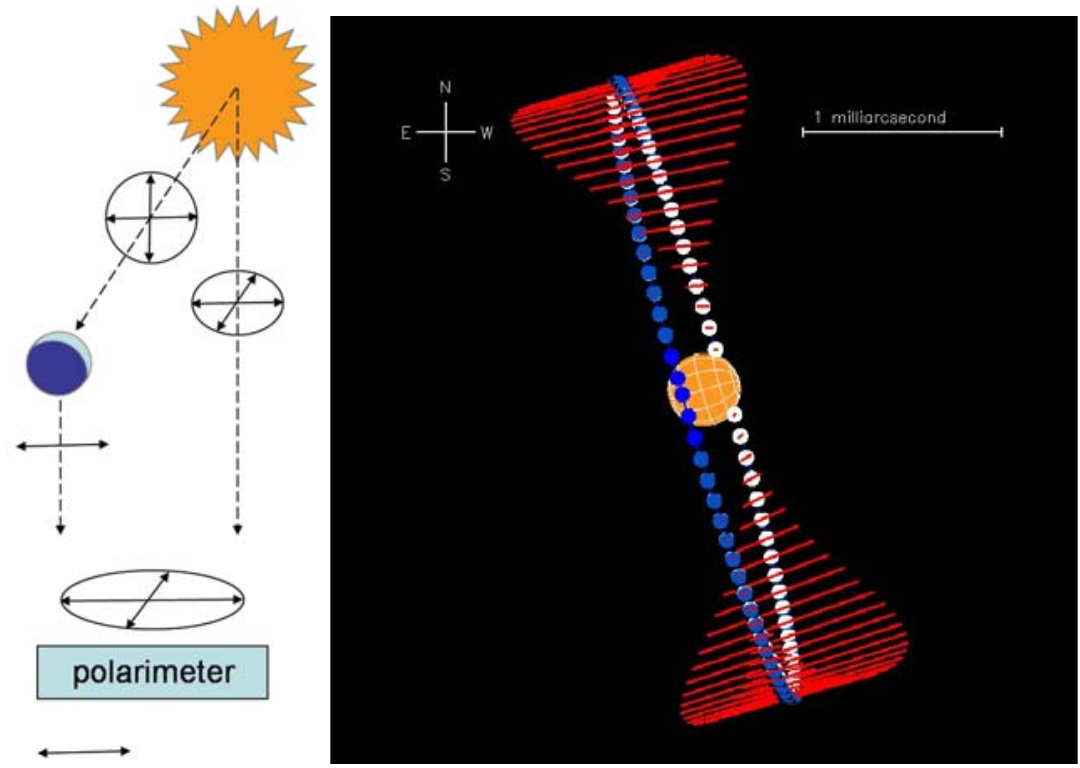}
\caption{{\it Left:} A star emits unpolarized light in all directions. The part of light that is scattered by a planet towards the observer is linearly polarized perpendicular to the scattering plane. Thus, the observer registers a partially polarized light from an unresolved star-planet system and can detect directly the scattered light using a polarimeter. 
{\it Right:} As the planet revolves around the star, the scattering angle changes and both  the direction and degree (red sticks) of polarization vary. For a circular orbit, as shown here for HD189733b, two peaks per orbital period can be observed near elongations. Therefore, polarimetry provides information on the planetary atmosphere outside transits.}
\label{fig:orbit}
\end{figure}

The physics of scattering in the planetary atmosphere is similar to that of the solar case. The light scattered off the planet is linearly polarized perpendicular to the scattering plane (Fig.~\ref{fig:orbit}, left). In general, when the planet revolves around the parent star, the scattering angle changes and the Stokes parameters vary. If the orbit is close to circular, two peaks per orbital period can be observed (Fig.~\ref{fig:orbit}, right). Thus, the observed polarization variability exhibits the orbital period of the planet and reveals the inclination, eccentricity, and orientation of the orbit as well as the nature of scattering particles in the planetary atmosphere. It is feasible therefore to detect the scattered light with polarimetry, even if an exoplanet is not spatially resolved, and hot Jupiters on close orbits are the best candidates for polarimetric detections. 

\begin{figure}[t]
\centering
\includegraphics[width=0.8\linewidth]{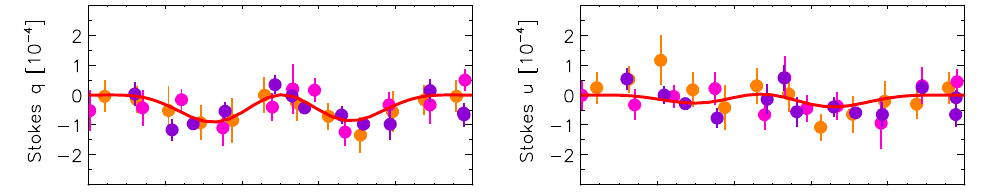}
\includegraphics[width=0.8\linewidth]{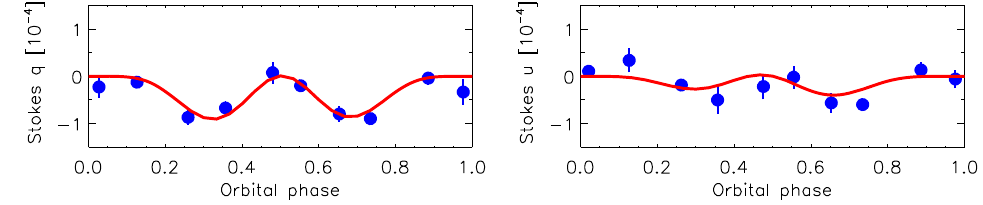}
\caption{Polarimetric data (normalized Stokes $q$ and $u$) for HD189733b in $U$ (pink) and $B$ (orange and purple for two different sets of data)  photometric bands (upper panels). 
The binned data are shown in lower panels (blue). The mean error of the binned data is 1.7$\cdot$10$^{-5}$. Curves are the best-fit solutions for a model atmosphere with Rayleigh scattering on dust particles. The standard deviation of the fit is 1.1$\cdot$10$^{-5}$.
}
\label{fig:hdstokes}
\end{figure}

Our first target was a very hot Jupiter HD189733b which is only 0.03\,AU away from the star. In 2006--2007 we observed the system in the $B$-band with the double image CCD polarimeter DiPol installed at the remotely controlled 60\,cm KVA telescope on La Palma \citep{Berdyuginaetal2008} and in 2008 in the $UBV$-bands with the TurPol polarimeter at the 2.5\,m Nordic Optical Telescope (NOT), La Palma \citep{Berdyuginaetal2010}. Our observations clearly reveal polarization peaks of 10$^{-4}$ near elongations as expected for a planet on a circular orbit (Fig.~\ref{fig:hdstokes}). 

\begin{figure}[!ht]
\centering
\includegraphics[width=0.4\linewidth]{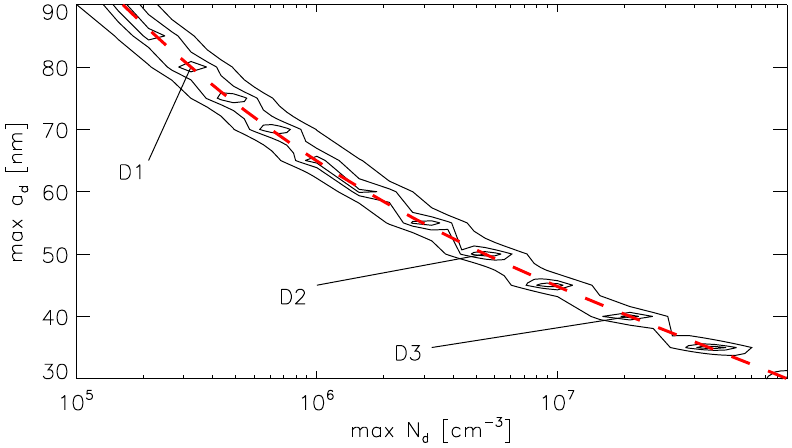}
\includegraphics[width=0.5\linewidth]{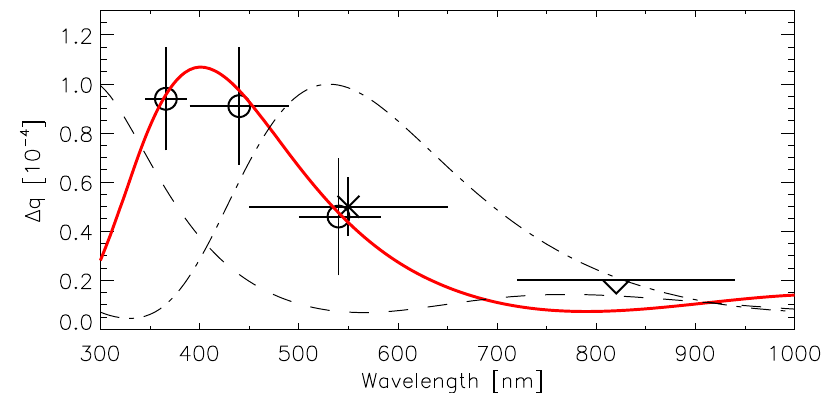}
\caption{{\it Left:} $\chi^2$-contours for peak values of the dust particle size $a_{\rm d}$ and number density $N_{\rm d}$. The confidence intervals of 68\%, 90\%, and 99\% are shown with solid lines. The dashed red line indicates a relation for acceptable solutions. 
{\it Right:} Stokes $q$ peak amplitudes in different passbands. Open circles are the $UBV$ amplitudes from \citep{Berdyuginaetal2010}, cross is the upper limit by \citet{Wiktorowicz2009}, and triangle is the upper limit for hot Jupiters by \citet{Lucasetal2009}. Horizontal bars indicate the passbands' FWHI. Solid red line is the polarization curve for the Model D2 with max $a_{\rm d}$=50\,nm and max $N_{\rm d}$= 5$\cdot$10$^6$\,cm$^{-3}$. Dashed and dashed-dotted curves are the same model but with max $a_{\rm d}$ of 40\,nm and 60\,nm, respectively. Our data are consistent with other measurements and indicate that polarization occurs due to Rayleigh scattering with a peak in the blue.}
\label{fig:hdmodel}
\end{figure}

To interpret the data, we constructed a semi-empirical model atmosphere where polarization is caused by Rayleigh scattering \citep{FluriBerdyugina2010}. From the phase-locked polarization variations we determined orbital parameters, including the positional angle on the sky plane (see Fig.~\ref{fig:orbit}, right). Our model atmosphere includes the following opacity sources: (i) scattering on H, H$_2$, He, CO, H$_2$O, CH$_4$, electrons, MgSiO$_3$ dust condensates, and (ii) continuum absorption due to free-free and bound-free transitions of H, H$^-$, H$_2^+$, H$_2^-$, He, He$^-$, Si, Mg, and Fe. To obtain the best fit to our data, we varied temperature, gas pressure, and dust particle size $a_{\rm d}$ and density $N_{\rm d}$ in the planetary atmosphere within the height range of 82,000--84,500\,km with a step of 100\,km. For each layer we calculated opacities and polarization and searched for the best-fit solution with a $\chi^2$ minimization procedure. The dust distribution was considered as a layer with a Gaussian profile, which is characterized by the peak values of $a_{\rm d}$ and $N_{\rm d}$. 
In particular, we found (in prep.) that there is a set of similar quality best-fit solutions with a relation between the peaks of $a_{\rm d}$ and $N_{\rm d}$ (Fig.~\ref{fig:hdmodel}, left).

The data taken at different wavelengths including upper limits by others clearly indicates the dominance of Rayleigh scattering in the optical and can be described by a single model atmosphere (Fig.~\ref{fig:hdmodel}, right). A transit curve calculated taking into account opacities in the atmosphere perfectly agrees with the transit data by \citet{Pontetal2007}. The geometrical albedo of our model atmosphere varies with wavelength and reaches 0.48 at 400\,nm, 0.2 at 550\,nm, 0.1 at 600\,nm, and 0.05 at 700\,nm, in agreement with upper limits for hot Jupiters. We conclude that the polarization in HD189733b occurs due to Rayleigh scattering in the atmosphere, most probably in on dust condensates of average size of about 20\,nm. 

Our results establish polarimetry as an important tool for studying directly exoplanetary atmospheres in the visible. In the near future it will be employed for non-transiting systems.

\section{Conclusions}\label{sec:conc}

This review demonstrates that polarimetry of cool atmospheres with the help of molecular physics is a maturing field with many theoretical tools developed and ready to be employed to uncover new insights into magnetic structures and inhomegeneities of cool atmospheres. In addition, as we learn about the physics of the processes responsible for polarization in molecular transitions, a great potential for interdisciplinary research is revealed. For instance, polarimetry of exoplanets and protoplanetary disks are new emerging fields where this expertise is of
great advantage.

\acknowledgements 
I am very grateful to all my collaborators and PhD students with whom it has been such a pleasure and also great fun to work together. We are all devastated that our friend Jean Arnaud had suddenly passed away at the time when we were preparing these proceedings.
This work during the past five years was supported by the EURYI Award from the ESF (www.est.org/euryi), the SNF grant PE002-104552, and the Academy of Finland grant 115417.


\end{document}